# A 130-MS/s 10-Bit Asynchronous SAR ADC with 55.2 dB SNDR

Ayan Mandal and Asish Koruprolu

***Abstract*—This paper presents a low-power 10-bit 130-MS/s successive approximation register (SAR) analog-to-digital converter (ADC) in 90 nm CMOS process. The proposed asynchronous ADC consists of a comparator, SAR logic block and two control blocks for the capacitive digital to analog converters (DAC). At a 1.2 V supply and 130 MS/s, the ADC achieves an SNDR of 55.2 dB and consumes 860 uW, resulting in a figure of merit (FOM) of 50.9 fJ/MHz. It achieves an ENOB of 8.8 bits with a differential input range of ±785 mV.**

## I. Introduction

Scaling in the gate length of CMOS devices has led successive approximation register (SAR) ADCs become a popular architecture in recent medium-to-high resolution and low-to-high speed applications. Applications of SAR ADCs range from biomedical instrumentation sensor platforms (low speed and ultra-low power) to emerging optical and data links (ultra-high speed). The primary advantages of SAR ADCs compared to other Nyquist architectures are lower power consumption, no high-gain amplifier stage requirement and small active area [1-2].

## II. System Design

One of the primary decisions to be made while designing a SAR ADC is choosing the full-scale range (FS). This critical decision defines the relative complexity of the Track and hold circuit compared to the comparator design and architecture. Choosing a low FS requires a stringent comparator design, as the least significant bit (LSB or $\Delta$ defined as $FS/2^B$, where B is the number of bits) is now smaller and therefore the regeneration constant must be higher for the same metastability probability. On the other hand, choosing a large full scale, leads to more non-linearities in the track and hold block, thus making its design complex. As a reasonable compromise the full-scale range was initially selected as 1.2 V. After several iterations the FS for the proposed design has been set to 1.6 V owing to the relatively complicated design of the comparator.

The following sections III- VII describes each block of the proposed SAR ADC design. The block diagram of the entire implemented architecture is shown in Figure 1.

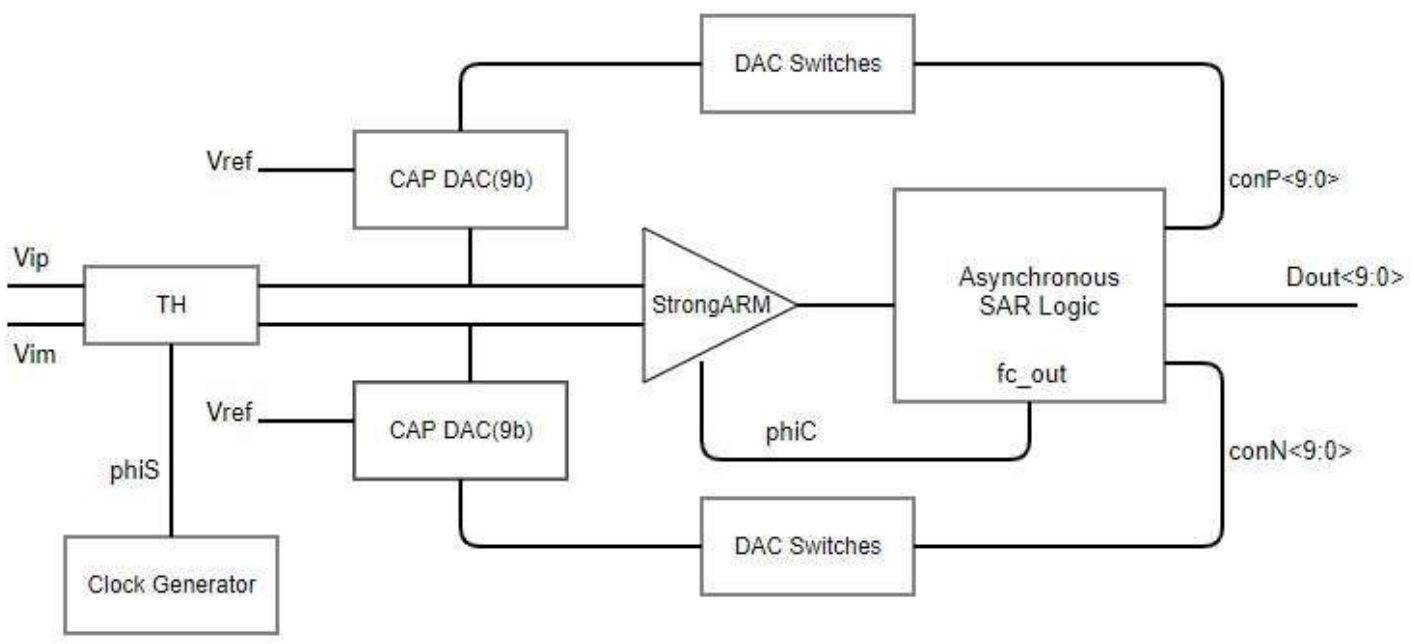

Figure 1: Block diagram of the proposed SAR ADC architecture.

## III. Track and Hold Circuit

### A. Architecture

The sampling switch in the track and hold (TH) circuit is an important part of the SAR ADC design. It is the major contributor of non-linearities in any SAR ADC system. It is also critical as it drives the capacitive DAC during the sampling phase and therefore must be fast enough to settle. Sizing the width of the switch (W) small can lead to large dynamic settling errors and introduce tracking non-linearities due to on-resistance variation. Whereas making W large can increase the power consumption and other non-linear effects such as charge injection and feedthrough. Therefore, a preferred switch architecture is one which minimizes non-linearities and can settle fast enough with a large DAC capacitance.

A single NMOS/PMOS transistor in the triode region can act as a switch and can provide fast settling but is not a good choice as the on-resistance depends on the input which leads to non-linearities. To achieve a more uniform on-resistance for the switch a transmission gate can be used but it doesn't give much benefit in terms of the tracking non-linearity. In-order to minimize the tracking non-linearities of the Track and Hold circuit, a Bootstrapped switch [3] has been used in this work. Well-designed bootstrapped switches can result in a 20-30 dB increase in the SNDR performance as compared to transmission gate topologies.

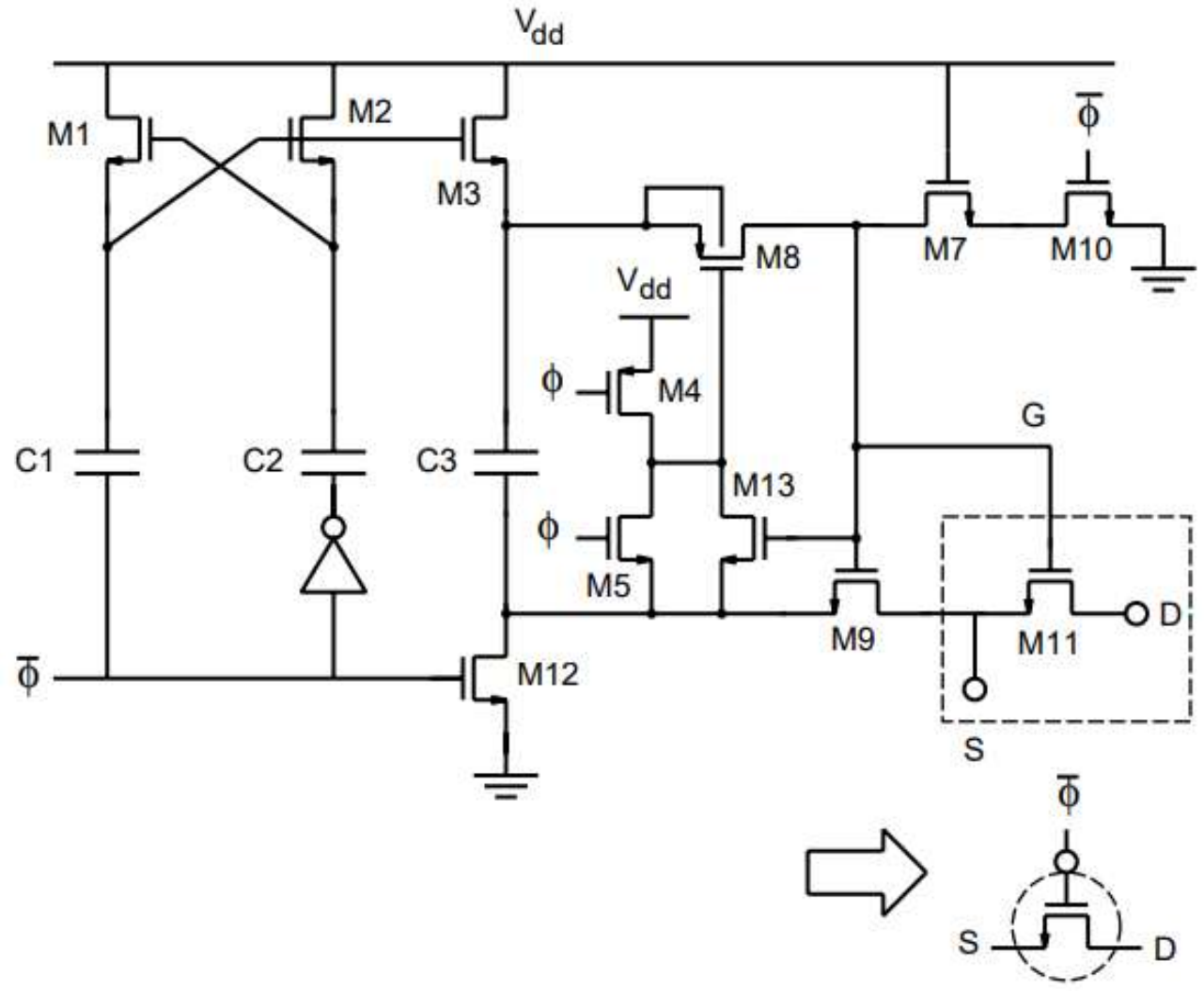

Figure 2: Bootstrap Switch transistor level circuit [3]

## B. *Equations and Trade-offs*

There are two main opposing factors governing the design of a bootstrap switch- a. the settling speed and tracking linearity (increase W of the switch transistor) and b. other non-linearities such as charge injection and feedthrough (decrease W of the switch transistor). The power consumed by other components in the ADC system is much larger than that of the bootstrap and therefore power is not a major factor in the design for this block. The tracking non-linearity of a bootstrap switch is related to the on-resistance of the bootstrap switch which varies as-

$$r_{on} = \frac{1}{\mu_n C_{ox}\left(\frac{W_n}{L_n}\right)\left(\frac{C_{boot}}{C_{boot}+C_{par}}V_{DD} - \frac{C_{par}}{C_{boot}+C_{par}}V_{in} - V_{tn}(V_{in})\right)} \quad (1)$$

The switch transistor should be large enough such that $Nr_{on}C_{DAC} = T_{track}$, which is the tracking time of the switch. N can be chosen to be between 5-10 depending on the required DAC settling. N is chosen to be around 9 for 0.01% settling. A choice of large N in the design leads to an extra noise penalty.

## C. *Design Strategy*

The sizing of each device in the bootstrap circuit (Fig. 2) is dependent on what purpose it serves in the circuit. A brief description of the design methodology is as shown below –

1. $C_{boot}$ strongly affects the non-linearity of the TH circuit as making it large reduces the dependence of Vin on $r_{on}$. We want to make $\frac{C_{par}}{C_{boot}+C_{par}} \approx 0$.
2. $M_3$ and $M_{12}$ need to charge $C_{boot}$ (which is a large capacitor) to $V_{DD}$ during the reset phase and therefore must be large enough. The fact that $C_{boot}$ does not discharge completely during the track phase and that the reset phase continues for 10 clock cycles relaxes the sizing constraint on $M_3$ and $M_{12}$.
3. $C_1$ must be sized large enough to boost the gate of NMOS $M_3$ to approximately $2V_{DD}$ (for $C_{boot}$ to have $V_{DD}$ across it during reset).
4. $M_7$ and $M_{10}$ pull down the gate of $M_{11}$ (switch transistor) to ground in the reset phase, which lasts for a long time and therefore they can be small.
5. $M_8$ and $M_9$ set the $V_{GS}$ of $M_{11}$ to $V_{DD}$ and therefore govern how quickly $M_{11}$ starts tracking. They are made moderately large so that $M_{11}$ can start tracking accurately.
6. $M_4$ and $M_5$ are responsible to pull up and pull down the drain of $M_{13}$, which is not a large transistor and therefore can be small sized.

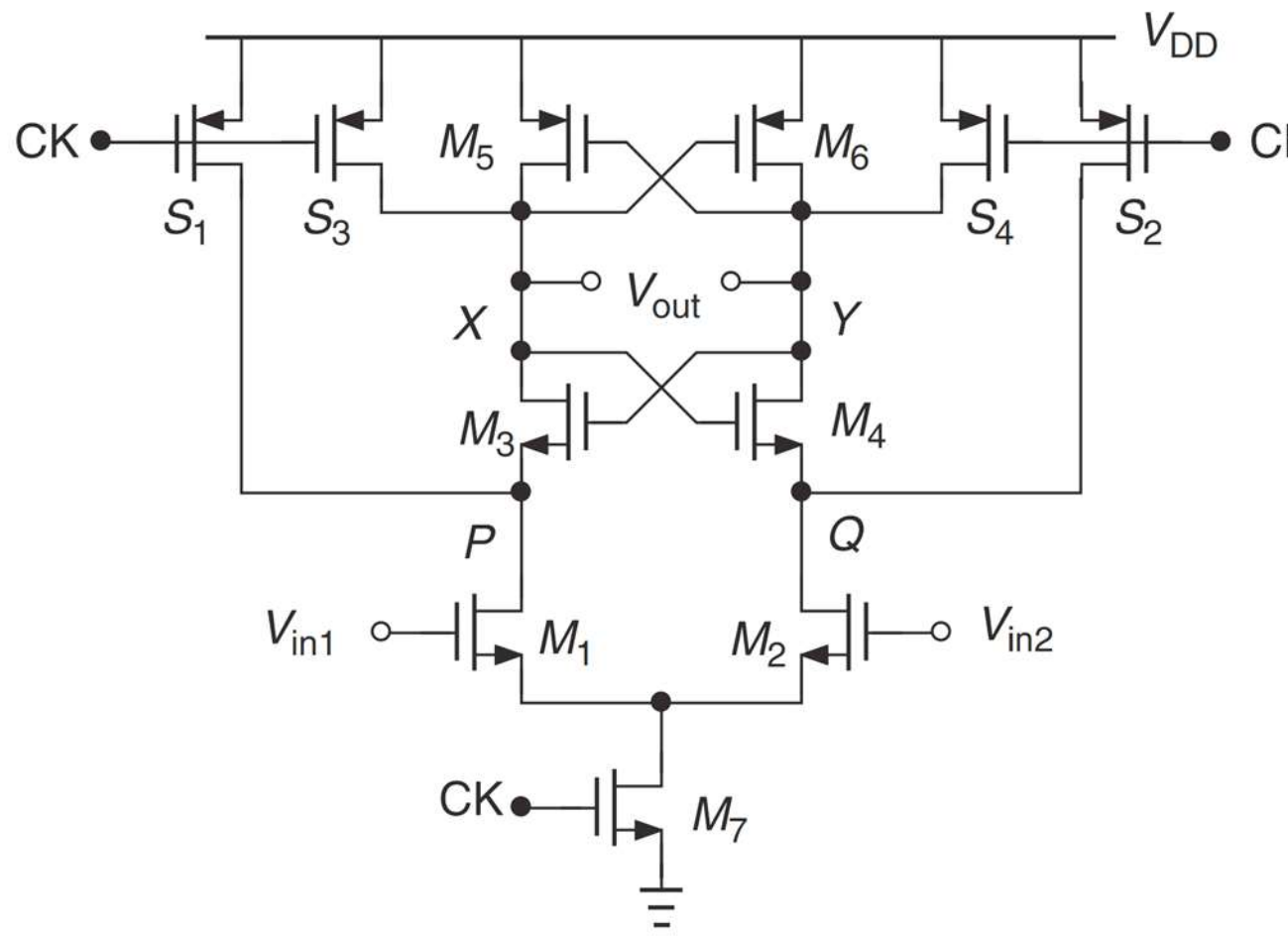

Figure 3: StrongARM Latch transistor level circuit [4]

# IV. Comparator Design

## A. *Architecture*

The comparator is another major block to focus on while designing a SAR ADC. A high sampling frequency ($f_s$) requires a fast comparator as the comparator must make B comparisons in one sampling cycle. Due to this the comparator is a major contributor to the power and noise in the system. Dynamic comparators, with clocked tail current sources have minimal static power consumption and are an obvious choice for low power designs. Keeping this in mind the StrongARM latch (Fig. 3) was chosen for this design. In addition to being low power, the StrongARM latch also produces a rail-to-rail output and has a low regeneration time owing to the inverter latches connected to the differential pair.

## B. *Equations and trade-offs*

The three main quantities defining the performance of a comparator are – power consumption, regeneration time constant and input referred noise. Due to the large signal behavior of comparator circuits it is difficult to come up with exact equations for these quantities. Although, approximate equations derived in [4] can be used to identify the major trade-offs.

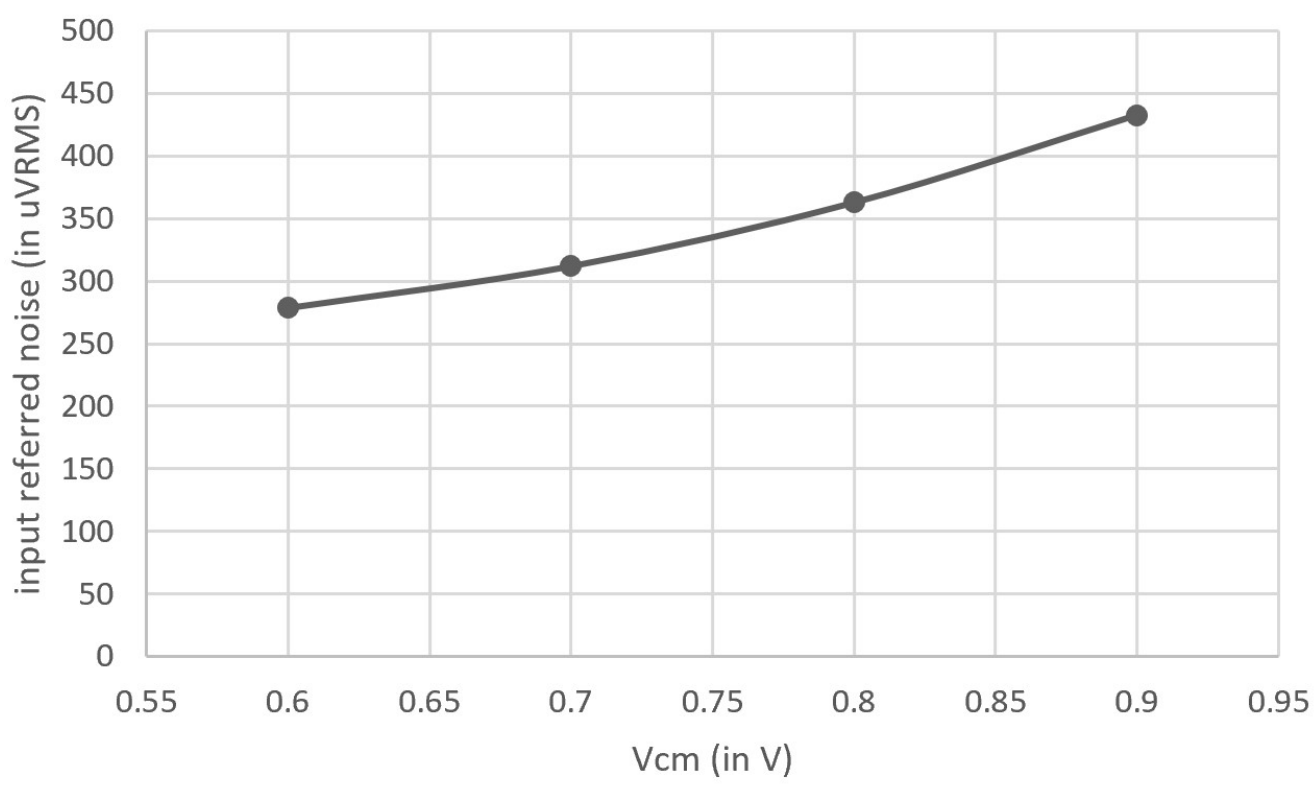

(a)

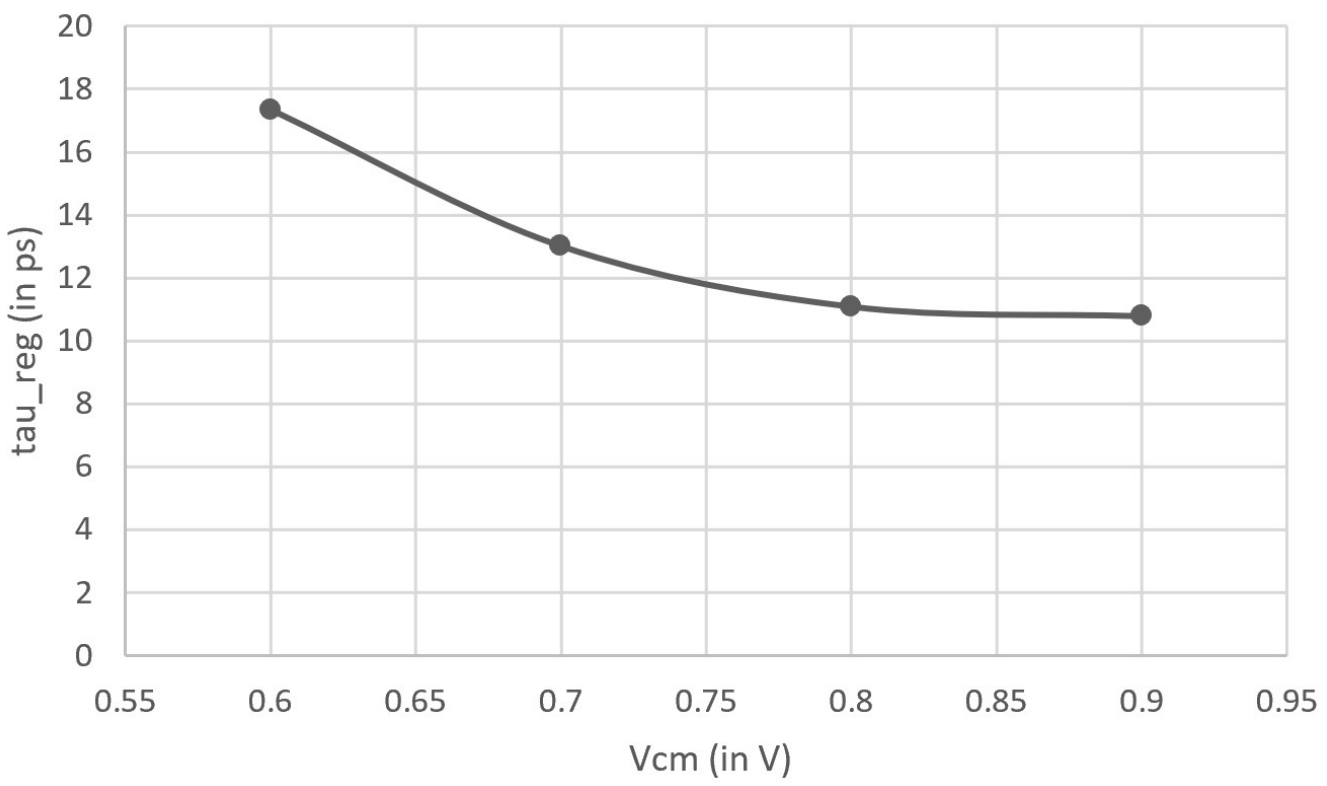

(b)

Figure 4: (a) Simulated input referred noise (b) $\tau_{reg}$ - with respect to variation in $V_{cm}$ of the StrongARM latch

$$V_{n,in}^2 = \frac{(V_{GS_{1,2}} - V_{THN_{1,2}})}{V_{THN}}\left[\frac{4kT\gamma}{C_{P,Q}} + \frac{(V_{GS_{1,2}} - V_{THN_{1,2}})}{V_{THN}} \cdot \frac{kT}{2C_{P,Q}}\right] \quad (2)$$

$$P = f_{ck}\left(2C_{P,Q} + C_{X,Y}\right)V_{DD}^2 \quad (3)$$

$$\tau_{reg} \approx \frac{C_{X,Y}}{g_{m5}} \quad (4)$$

It is clear from (2) and (3) that the input referred noise ($V^2_{n,in}$) of the comparator can be reduced by either reducing the common mode voltage (reduce $V_{GS}$) or increasing the capacitances $C_{P,Q}$ or $C_{X,Y}$ (increasing size of the Latch transistors). As seen in Figure 4, this clearly trades-off with both the power consumption and the speed of the comparator. Therefore, in this design the noise performance is designed to be just enough to guarantee an ENOB of 8.8 bits, which allows a lower capacitance value and higher frequency of operation which improves the figure of merit (FOM = $P_{tot}/f_s^2$).

### C. Design Strategy

In order to do quick iterations on the comparator design a testbench was set up for the block to measure the input referred noise (using pnoise analysis in Cadence Virtuoso) and the regeneration time constant. The comparator was designed to have a very low regeneration constant (4) of 13 ps, as this would guarantee functionality at high frequencies [5]. The first step in designing the comparator was calculating the noise budget for the comparator based on the target SNDR specification and other noise sources, which can be done using (5).

$$V_{n,in,comp}^2 + \frac{2kT}{C_{DAC}} + V_{n,quant}^2 + V_{n,TH,non-linearity}^2 = \frac{P_{sig}}{10^{\frac{SNDR}{10}}} \quad (5)$$

To get a good balance of speed and power consumption, the comparator transistors are made to operate in moderate-strong inversion (for the same $g_m$, capacitances are higher in weak inversion). The transistors have been sized keeping this operating regime in mind. Although, the comparator operating regime changes during comparisions, the initial operating region for the input differential pair sets the initial slew rate and the gain accrued before regeneration, which can simplify the design of the comparator to some extent. The top latch acts like an inverter and is sized greater than 2:1 (PMOS/NMOS width ratio), as increasing its transconductance reduces the $\tau_{reg}$.

Following the above design strategy, the comparator was designed to operate at a common mode voltage of 700mV with 13 ps regeneration time constant and 312 $\mu V_{rms}$ input referred noise.

## V. DAC switches

### A. Architecture

The DAC switches although not critical to the performance, can impact the SNDR of the overall ADC significantly if not designed carefully. The DAC switches need to be sized sufficiently large for the CAP DAC to settle with reasonable accuracy when the comparator is off. Since the input to the DAC switches is constant (either $V_{REF}$ or ground), linearity is not an issue here. $V_{REF}$ is chosen as FS/2. Additionally, since one switch either pulls down to zero of pulls up to $V_{REF}$ we can use NMOS for the former and PMOS for the later.

### B. Equations and trade-offs

The major trade-offs concerning the DAC switches are with respect to the power consumption and CAP DAC settling time. The switches connected to the MSB of the CAP DAC can be large and therefore must be sized carefully to achieve settling while avoid unnecessary power consumption. Sizing them larger improves settling time but will drastically increase the power consumption (and other non-linearities such as charge injection). The on-resistance of the switch should be designed according to (6), where $r_{on,i}$ is the on-resistance of the $i^{th}$ switch, $C_{DAC,I}$ is the DAC capacitance of the $i^{th}$ bit and $t_{\phi c=low}$ is the off-time of the generated comparator clock using the SAR logic.

$$r_{on,i} = \frac{1}{N\, C_{DAC,i}\, t_{\phi c=low}} \quad (6)$$

### C. Design Strategy

Using the above strategy, it is possible to size the switches such that larger capacitances are driven with larger switches such that $r_{on}C$ is the same for all the switches. The on-resistances of the switches can be determined using simulation and then assuming a linearly inverse relationship between W and $r_{on}$ the required size of each switch is estimated.

## VI. SAR LOGIC

### A. Architecture

Synchronous and asynchronous are the most commonly used schemes to design the logic that controls the DAC switching. In the synchronous architecture, one clock period ($T_{CLK}$) is allocated for every bit cycle of the successive approximation algorithm. The prime difference in asynchronous logic is the use of different clock cycle duration for each bit. SAR logic performs a comparison only after the comparator output settles. Implementation of a separate clock generator for the comparator is avoided as asynchronous logic generates its own clock to drive the comparator. Asynchronous logic enables an improvement in the metastability rate for a given number of allotted regeneration time constants (N) per bit comparison.

### B. Equations and trade-offs

The metastability rate target in the proposed model is $10^{-7}$, which gives an upper bound estimate of the longest it takes for the comparator to settle. This is termed $T_{hard}$ which is given by (7) where $V_{DD}$ is the supply voltage, $P_{meta}$ is the metastability rate, $\Delta$ is the least significant bit (LSB) value and $A_v$ is the gain of the comparator stage right before the regeneration phase starts [6]. The maximum achievable sampling rate ($f_{s|max}$) is given by (8), where $T_{delay}$ is the logical delay in generating the asynchronous comparator clock, $T_{easy}$ is the time taken for all the comparisons to settle expect the hard comparison. $T_{easy}$ for a 10-bit SAR ADC is roughly equal to 39 $\tau_{reg}$ [6].

$$T_{hard} = \tau \ln\left(\frac{2\,V_{DD}}{A_v\ P_{meta}\,\Delta}\right) \tag{7}$$

$$\frac{1}{f_{s\,|\,max}} < T_{easy} + T_{hard} + (B-1)\,T_{fix} + B\,T_{delay} + T_{track} \tag{8}$$

### C. Design Strategy

The asynchronous logic in this proposed model has been implemented in VerilogA. Each input to the SAR block is modeled with 50 fF capacitance and each output feeds into a buffer that is sized according to the load that is being driven. Implemented asynchronous logic resulted in a 30% boost in the sampling frequency for the SNDR (signal-to-noise distortion ratio), which improves (lowers) our figure of merit.

## VII. DIGITAL TO ANALOG CONVERTER (DAC)

### A. Architecture

Preserving the common-mode while the DAC switches is important as the Strongarm latch comparator is very sensitive to changes in $V_{cm}$. The DAC in the proposed circuit is split into two halves and the switching is performed in such a way that the voltage change on the differential terminals of the comparator are equal and opposite. Hence, the common-mode is preserved while the differential voltage keeps changing in every iteration. Monotonic switching helps reduce power consumed by the DAC block.

### B. Equations and trade-offs

The major trade-off associated with the DAC is the value of the unit capacitance used in the binary-weighted DAC. Lower net capacitance of the DAC consumes less power but results in higher kT/C sampling noise.

The net input differential full scale range ($V_{FS|net}$) is determined by (9) due to parasitic capacitances ($C_p$ which also includes the comparator's input capacitance). $V_{FS}$ used here is $2V_{REF}$ (as defined previously).

$$V_{FS\,|\,net} = V_{FS}\,\frac{C_{DAC}}{C_{DAC} + C_p} \tag{9}$$

### C. Design Strategy

The minimum possible unit capacitance of 2.5 fF has been used as the sampling noise isn't the dominant noise contributor in the proposed design. Metal insulator metal capacitors ($C_{mim}$) have been used in implementing the DAC and the thin plate of each capacitor has been connected to the input of the comparator to reduce the effect of parasitics. The total DAC capacitance as seen at the output of the bootstrap circuit block is 1.3 pF and the parasitic contribution 20 fF seen at the node results in $V_{FS|net}$ = 785 mV.

A variant of implementing the DAC more commonly known as a Split-DAC has also been explored [7]. The split-DAC reduces the sizes of the capacitors that are required as compared to a binary-weighted DAC, by using an attenuation capacitor ($C_{att}$) in between the LSB and the MSB DAC capacitors. The hybrid common-mode preserving split-DAC designed wasn't chosen for this implementation due to the large non-linearity (due to parasitic capacitances) and the increase in the corresponding sampling kT/C noise due to the reduction of the capacitance by 16 times, although the corresponding power saving in this block is 37.5% [8].

TABLE I
ADC SPECIFICATION

| Parameter | Quantity |
|---|---|
| Process | 90 nm |
| Supply Voltage | 1.2 V |
| Power Consumption | 860 μW |
| Figure of Merit | 50.9 fJ/MHz |
| Differential Input Range | ± 750 mV |
| SNDR | 55.2 dB |
| SFDR | 60.6 dB |
| ENOB | 8.8 bits |

## VIII. Results

The proposed 10-bit SAR ADC is simulated using 90-nm CMOS models to characterize the power and noise performance. The achieved signal to noise distortion ratio (SNDR) for an input sinusoid with 750 mV amplitude and (3/64)fs is 56.4 dB (Fig. 5). To validate functionality, the system was tested at the Nyquist frequency and can achieve 55.2 dB SNDR (Fig. 6). In addition, to characterize the figure of merit (FOM), the total power consumption of the system was measured. A block level power consumption distribution is shown in Fig. 7. A more detailed description of the ADC specifications can be found in Table 1.

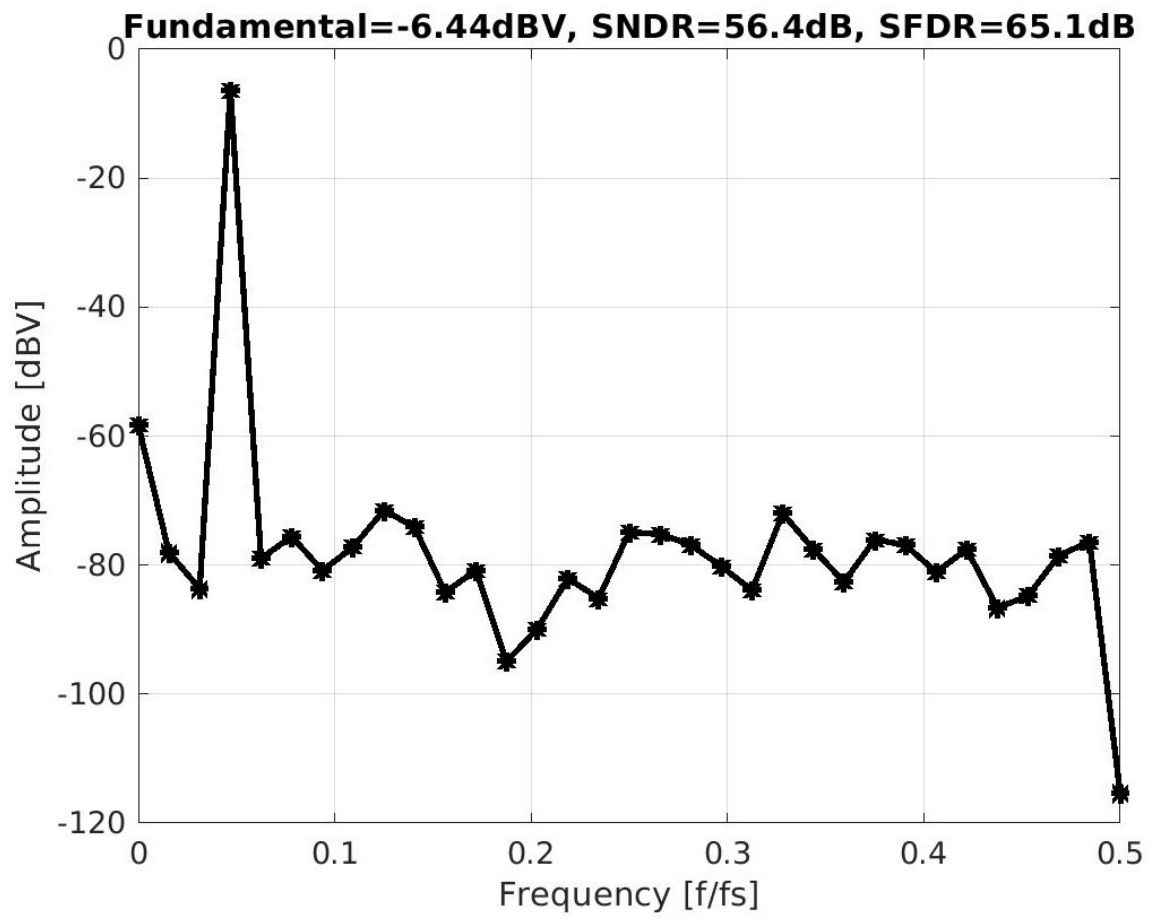

Figure 5: 64-point FFT of the ADC output for an input sinusoid signal of (3/64)$f_s$

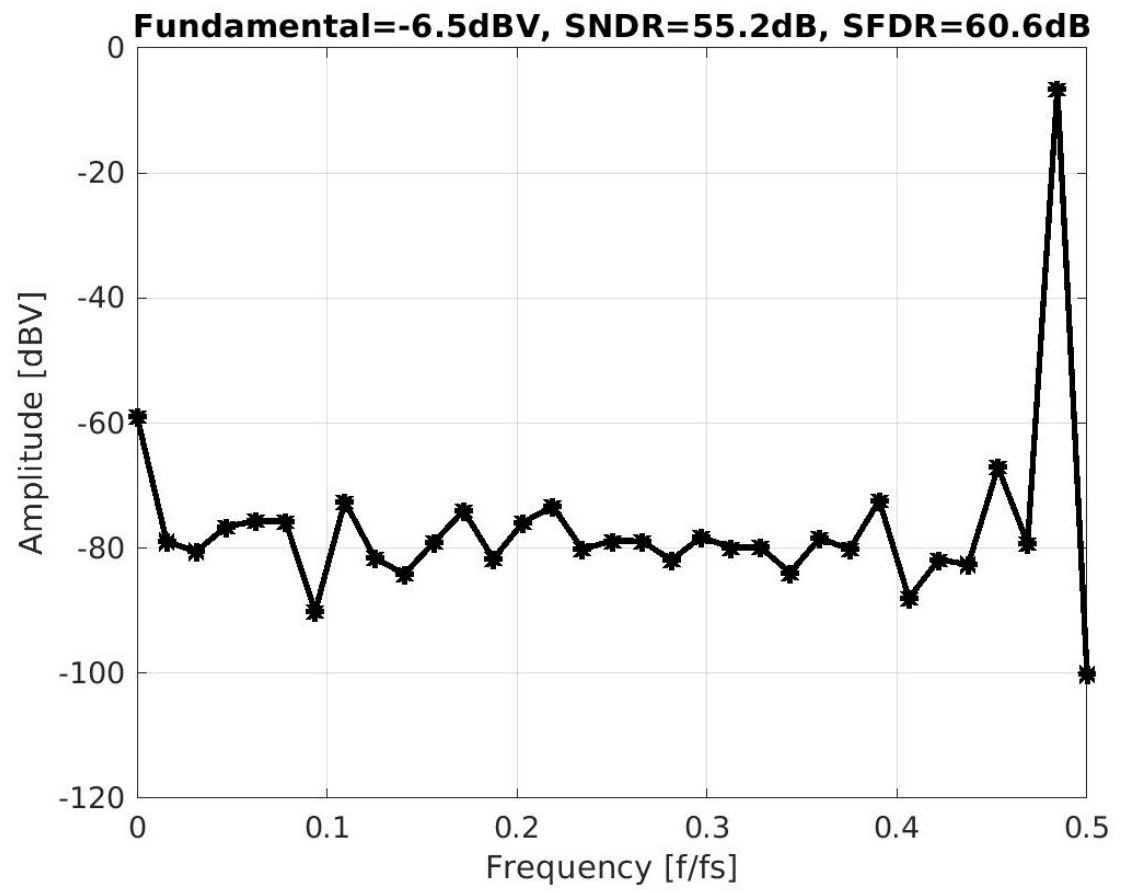

Figure 6: 64-point FFT of the ADC output for an input sinusoid signal of (31/64)$f_s$ with peak-to-peak amplitude of 1.5V

## IX. Conclusion

A low-power 10-bit 130-MS/s SAR ADC designed on a 90-nm CMOS process has been presented. The system architecture is described, and the trade-offs taken into consideration while designing the SAR ADC are highlighted in this paper.

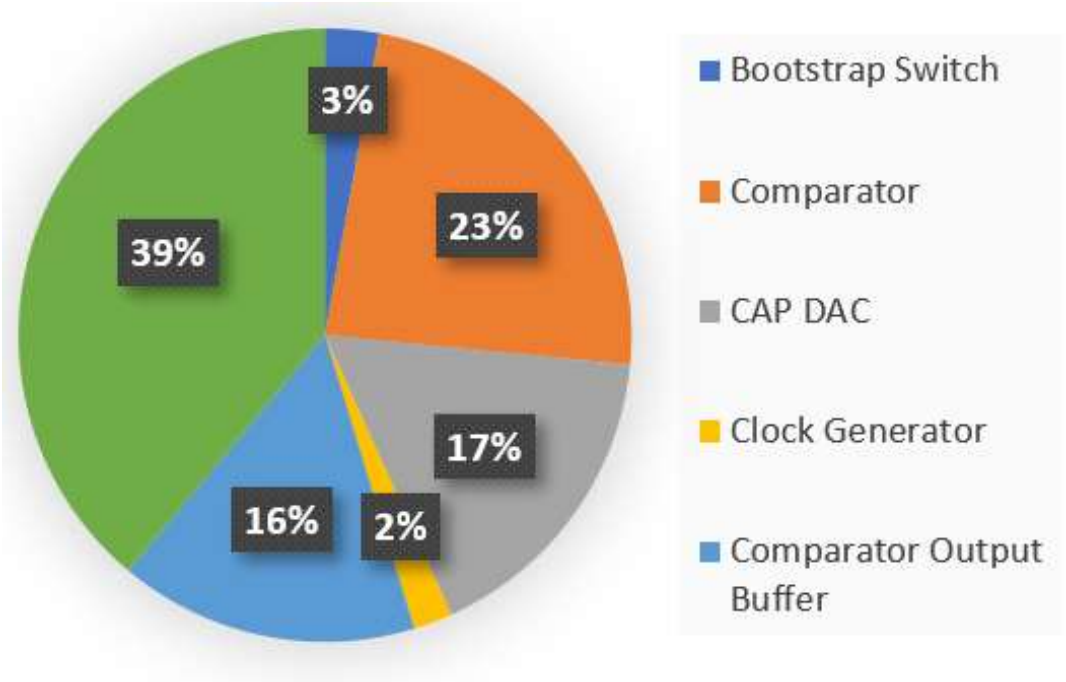

Figure 7: Pie-chart comparing the power consumed by the various blocks in the SAR ADC design